\documentstyle[12pt,aaspp4]{article}

\lefthead{von Braun, Chiboucas, and Hurley-Keller}
\righthead{Observational Mishaps: a Database}

\begin{document}

\title{Observational Mishaps: a Database}

\author{Kaspar von Braun, Kristin Chiboucas, Denise Hurley-Keller}

\affil{University of Michigan}
\authoraddr{Department of Astronomy, Univ. of Michigan, Ann Arbor, MI 48109-1090}
\authoremail{kaspar@astro.lsa.umich.edu, kristin@astro.lsa.umich.edu, 
denise@astro.lsa.umich.edu}

\begin{abstract}

We present a World-Wide-Web-accessible database of astronomical images
which suffer from a variety of observational problems, ranging from
common occurences, such as dust grains on filters and/or the dewar
window, to more exotic phenomena, such as loss of primary mirror 
support due to the deflation of the support airbags. Apart from
its educational usefulness,
the purpose of this database is to assist astronomers in
diagnosing and treating errant images {\it at the telescope}, thus
saving valuable telescope time. Every observational mishap contained
in this on-line catalog is presented in the form of a GIF image, a
brief explanation of the problem, and, when possible, a suggestion for
improving the image quality.

\end{abstract}


\section{Introduction}

It is not uncommon for an astronomical image obtained after a lengthy
integration to reveal that all is not well.  As a consequence,
telescope time is sacrificed identifying the problem.  In an effort to
shorten this investigation period, we have created a catalog of
astronomical images bearing signatures of a range of mishaps
encountered during observing runs. Included with each image is an
explanation of the cause of the problem as well as a suggested
solution.  Since a large number of observatories today are connected
to the Internet, the World Wide Web (WWW) was chosen as the ideal
medium for presenting this collection of images.

Initially, the purpose of such a collection was to assist
new graduate student observers at Michigan-Dartmouth-MIT (MDM)
Observatory who frequently observe without the
benefit of a more experienced observer.  The aim was to provide these
students with a means of quickly pinpointing the 
underlying problem affecting the image quality. 
This idea grew into a WWW accessible
database complete with explanations of the ``mishaps'' responsible 
for the deterioration of the images, as well as suggested solutions.

\section{The Format of the Database}

\begin{figure}
\plotone{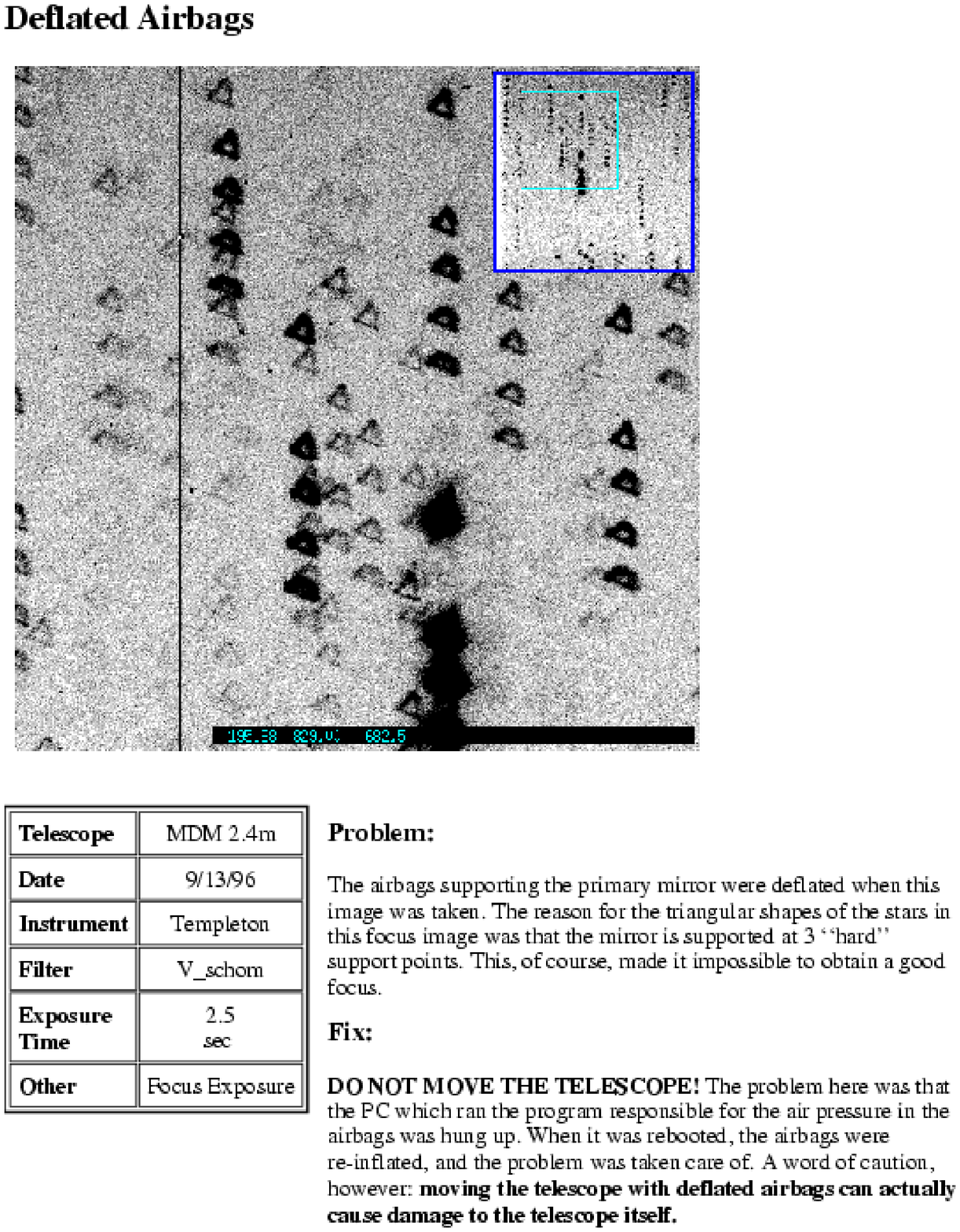}
\caption{ 
A typical webpage in the database.  \label{fig1.small}}
\end{figure}

Every WWW page in this catalog contains an inverted colormap GIF image
of the mishap, a table listing relevant information about the image
(telescope, date, instrument, filter, exposure time), a brief
description of the problem, and, if available, a suggestion of how to
fix it.
In a few cases, the cause of the problem could not be
determined. These were dubbed ``Unsolved Mysteries'', and no
explanation of the problem or suggestion for a fix are given.  
Since it is possible for one 
problem to manifest itself in a variety of ways, 
multiple images of the same mishap are presented where appropriate,
cross referenced with the help of hypertext links.  
For example, condensation on the dewar window can appear as a filamentary 
structure or as a bright extended feature with cusps, depending
on the locations of light sources in the field of view. 
For the more common problems of 
astigmatism, coma, bad guiding/focusing, and poor seeing, we
have provided supporting plots/images where applicable
via links on the relevant pages.  
Examples include radial profile plots across a
stellar image or multiple images of the same field taken in different
seeing conditions.
In Fig.~\ref{fig1.small}, we show an example of a typical page in 
the database, along with the explanation of the problem and a suggestion
for the solution.

\section{The Structure of the Database}

Much consideration was given to effectively structuring the image
catalog.  Rather than sorting the images by cause, which is probably
unknown to the astronomer accessing the database, we have grouped them
by symptom.  We provide the following two options for searching the
database:

\begin{enumerate}

\item The user may browse the complete list of compiled images.
This list features links to the various mishap pages as well as a
brief description (1 - 2 lines) of the 
symptoms in the corresponding image.

\item The other option is to first broadly classify the image
based on its symptoms and then choose the appropriate web page from a
smaller list. This option will likely be more practical with an
increasing number of images in the database.  Apart from the
frequently occuring problems of bad seeing/focusing/guiding, fringing,
dust rings, and reflections, the current revision of the database
lists the following as the top categories:

\begin{itemize}

\item Unusual Appearance of Objects in the Image: familiar
objects in the image, such as galaxies, stars, etc, have an unexpected
appearance (e.g., guider jumps, deflated airbags, etc).

\item CCD and Electronics Features: 
features seem to be correlated with the CCD rows or columns, or they
are otherwise suspiciously electronic in appearance 
(e.g., readout errors, shutter failure, etc).

\item Unexpected Objects in the Image and other External Interference:
unexpected features obviously not due to the CCD or the electronics
appear in the image (e.g.,
occulting dropout shutter, condensation on the dewar window, etc).

\item Unsolved Mysteries: as mentioned above, these are the cases for
which we have so far not been able to determine the cause of the
problem.

\end{itemize}

Each of the above links leads to a list of mishap pages in that
category with a brief description of the corresponding image
appearance.

\end{enumerate}

\section{The Location of the Database}

The Observational Mishaps Database can be accessed at

\noindent
http:$//$www.astro.lsa.umich.edu$/$mishaps$/$mishaps.html.

It is also directly accessible from the University of Michigan
Astronomy Department Home Page, whose URL is

\noindent
http:$//$www.astro.lsa.umich.edu.

\section{Additional Remarks}

We have created a database of images which are
deteriorated by the effects of various mishaps encountered during
astronomical observing runs. Its structure was designed to help users
quickly identify the cause of the poor image quality, thus saving
telescope time.  In addition to being widely accessible via the WWW,
the advantage of such an on-line catalog is its versatility. Unlike a
printed catalog, the on-line version can very easily be updated,
corrected, and expanded, so that everytime the database is accessed
the user will find it in its most up-to-date form.

Due to the practically infinite number of possible problems during
observing runs, this collection is clearly far from and 
impossible to complete.
Its usefulness, however, is obviously directly related 
to the number of examples
it contains, and therefore we would appreciate any
contributions by the astronomical community in the form of examples
which might fit into this collection.  Instructions for the submission
of such images are given in the database.

Furthermore, we realize that some of our interpretations of the
mishaps, as well as some of our suggestions on how to improve the
images, may be incorrect or incomplete.  While it is our intention 
to regularly update and improve this database, we welcome
any input about the database in general, its structure, or even
individual examples.

\acknowledgments

We would like to express our gratitude to the following people who
contributed to this project by supplying examples and/or providing
explanations of some of the mishaps: Gary Bernstein, Mario Mateo, 
Eric Miller, Patricia 
Knezek, Kelly Holley-Bockelmann, Lynne Allen, Michel Festou, and 
Doug Welch.

 
 
\end{document}